\documentstyle[12pt]{article} \topmargin=0.05in
\begin{document}
\def\t{\times}\def\p{\phi}\def\P{\Phi}\def\a{\alpha}\def\e{\varepsilon}
\def\be{\begin{equation}}\def\ee{\end{equation}}\def\l{\label}
\def\0{\setcounter{equation}{0}}\def\b{\beta}\def\S{\Sigma}\def\C{\cite}
\def\r{\ref}\def\ba{\begin{eqnarray}}\def\ea{\end{eqnarray}}\def\n{\nonumber}
\def\R{\rho}\def\X{\Xi}\def\x{\xi}\def\La{\Lambda}\def\la{\lambda}
\def\d{\delta}\def\s{\sigma}\def\f{\frac}\def\D{\Delta}\def\pa{\partial}
\def\Th{\Theta}\def\o{\omega}\def\O{\Omega}\def\th{\theta}\def\ga{\gamma}
\def\Ga{\Gamma}\def\h{\hat}\def\rar{\rightarrow}\def\vp{\varphi}
\def\inf{\infty}\def\le{\left}\def\ri{\right}\def\foot{\footnote}
\def\un{\underline}\def\ve{\varepsilon}\def\po{\propto}

\begin{center}

{\Large\bf Quantisation of Extended Objects}
\vskip 0.5cm
{\large\it J.Manjavidze\foot{Permanent address: Inst. Phys., Tbilisi,
Geogia}, A.Sissakian}\\ JINR, Dubna, Russia
\end{center}
\vskip 1cm
\begin{abstract}

The `strong-coupling' perturbation theory over the inverce
interaction constant $1/g$ near the nontrivial solution of Lagrange
equation is formulated. The ordinary `week-coupling' perturbation
theory over $g$ is described also to compare both perturbation
theories.  The `strong-coupling' perturbation theory is developed by
unitary mapping of the quantum dynamics into the space with local
coordinates of (action, angle)-type.

\end{abstract}
\section{Introduction}\0

The extended (soliton-like) objects quantisation problem considered
in this paper have more than twenty years old history, see the review
papers \C{jack, nev}. But absence of progress in this field evokes
anxiety, noting a number of unsolved by this reason important physical
problems.

One of possible  solutions of this problem was offered in
\C{hep}. The  aim of this article is  to show that the approach
described in this paper leads to the strong coupling perturbation
theory over inverse interaction constant $1/g$.

Our approach is based on the idea \C{yaph} that the
measure $DM$ of the functional integral representation for
\be
\R(E)=\int du_1du_2|<u_2;E|u_1;E>|^2
\l{1.1}\ee
is $\d$-like (Diracian):
\be
DM(u)=\prod_{x,t}du(x,t)\d\le(\f{\d S(u)}{\d u(x,t)}-j(x,t)\ri),
\l{1.2}\ee
where $j(x,t)$ is the random force of quantum excitations. We will
consider the symplest one particle quantum problem and the states in
(\r{1.1}) are described by the boundary values of coordinate $u_i$
and energy $E$.  In Sec.2.1 the physical basis of (\r{1.2}) will be
discussed and full derivation will be given in Sec.2.2.

The $\d$-function of (\r{1.2}) means that the strict space-time
local equality:
\be
-\f{\d S(u)}{\d u(x,t)}=j(x,t)
\l{1.3}\ee
defines the complete set of necessary contributions. The general
properties of theory defined on the $\d$-like measure is listed in
Sec.2.3. Note, (\r{1.3}) is not the consequence of Hamiltonian
variational principle and (\r{1.2}) will be derived in Sec.2.2,
proceeding from the conservation of total probability (unitarity
condition) \C{yaph}.

Eq.(\r{1.3}) shows that a transformation of kinetic part of Lagrangian
without fail induce the tangent transformation  of quantum source
$j(x,t)$ \C{gul}.  We  will use  this possibility  to  describe the
quantum dynamics  in useful terms.   Namely, we will apply the
canonical transformation of (\r{1.2}) to the  collective
coordinates. They will have  a meaning of (action, angle)-type
variables and will form  the cotangent foliation $T^*\O$ to the
incident phase space $\O$. So, used in this paper transformation is
the ordinary momentum mapping of classical mechanics \C{abr}:
\be
J:~(u,p)\to (\x,\eta),
\l{1.4}\ee
where $p$ is the conjugate to  $u$ momentum, see Secs.4.1 and 4.2.
The Hamiltonian description will be useful by this reason.

In this paper we restrict ourselves by one dimensional quantum
mechanics assuming that
\be
v(u;g)=\f{1}{g}v(g^{1/2}u;1)\equiv \f{1}{g}v(g^{1/2}u),
\l{1.5}\ee
where, for simplicity, $v(u)$ is the potential hole with one minimum
at $u=0$ and $g$ is the interaction constant. Then the nontrivial solution
$u_c$ of eq.(\r{1.3}) would be singular at $g=0$:
\be
u_c=O(g^{-1/2}).
\l{1.6}\ee
in the lowest order over $j$.

We will find that the transformed perturbation theory presents the
expansion over $\sim 1/g$ and the expansion coefficients are simply
calculable iff $T^*\O$ is the homogeneous and isotropic space

This result cardinally distinguished from the week coupling
perturbation theory developed in \C{jack, nev, jew}. We will derive
in Sec.3  this ordinary perturbation series over $g$ using the measure
(\r{1.3}) to show exactly where we turn from habitual way to formulate
new perturbation theory. Strictly speaking, there is not any
connections among both perturbation theories and they are dual to each
other, see Sec.5.

The paper is organised as follows. In Sec.2 we will find the integral
representation for $\R$ with measure (\r{1.2}). In Sec.3 we will
describe   the week-coupling perturbation theory. In Sec.4 the mapping
(\r{1.4}) for  the quantum system is described   to show the
decomposition over $1/g$ and the rule of calculation of corresponding
coefficient will be given. In concluding Sec.5 we will offer the
dynamical interpretation of new perturbation theory.

\section{Unitary definition of the functional measure}\0

Starting this section we will try to explain the role of unitarity in
definition of functional measure, Sec.2.1. Then we will show as the
d'Alembert's variational principle may be $derived$ for quantum
systems (Sec.2.2) and, at the end, the general properties of theory on
the $\d$-like measure will be offered.

\subsection{Formulation of method}

To calculate the bound state energies $E_n$ it is enough to consider the
trace\C{dash}:
\be
R(E)=\sum_n\int du \f{\psi_n(u)\psi^*_n(u)}{E-E_n-i\e}=
\sum_n\f{1}{E-E_n-i\e}
={\rm Sp}~\f{1}{E-{\bf H}-i\e},
\l{2.1}\ee
where ${\bf H}$ is the Hamiltonian operator and $\e\to+0$ and
the wave functions $\psi_n$ ortho-normalizability was used.

The semiclassical approximation leads to
\be
R(E)\sim \sum^{\infty}_{k=0}e^{ik(S_1(u_c)-\pi)}=
\f{1}{1+e^{iS_1(u_c)}},
\l{2.1'}\ee
where $S_1(u_c)$ is the action on the elementary (one period) closed path
trajectory $u_c=u_c(E)$ \C{dash}. The position of poles in (\r{2.1'})
defines the value of $E_n$.

Note now that
$$
\f{1}{E-E_n-i\e}={\rm P}\f{1}{E-E_n}+i\pi\d(E-E_n)~{\rm at}~\e=0,
$$
i.e. it is not necessary to calculate the real part since it did not
contain the masurable value of energy $E_n$:
$$
{\rm P}\f{1}{E-E_n}=0~{\rm при}~E=E_n.
$$
By this reason, following to \C{yaph}, we will calculate much more
simple quantity
\ba
\e\R(E)=\e\sum_{n_1,n_2}\int dx_1 dx_2 \f{\psi_{n_1}(x_1)
\psi^*_{n_1}(x_2)}{E-E_{n_1}-i\e}
 \f{\psi^*_{n_2}(x_1)\psi_{n_2}(x_2)}{E-E_{n_2}+i\e}=
\n\\
=\e\sum_n\le|\f{1}{E-E_n-i\e}\ri|^2=\f{1}{2i}\sum_n\le\{
\f{1}{E-E_n -i\e}-\f{1}{E-E_n +i\e}\ri\}=
\n\\
=\pi\sum_n\d(E-E_n)=\pi~{\rm Sp}~\d(E-{\bf H})={\rm Im}~R(E).
\l{2.2}\ea
Therefore, we wish exclude from consideration the unnecessary
contributions\foot{`Unnecessary' means for us the unmeasurable
quantity in {\it given} experiment.} with $E\neq E_n$. It should be
noted that we exclude continuum of contributions contained in ${\rm
Re}\{{1}/{E-E_n-i\e}\}$ and leave the set of point-like contributions
${\rm Im}\{{1}/{E-E_n-i\e}\}$, but with infitite amplitudes.

We can find \C{yaph} that
\be
\R(E)\sim\sum^{+\infty}_{k=-\infty}e^{ik(S_1(u_c)-\pi)}=
2\pi\sum_n\d(S_1(u_c)-(2n+1)\pi)
\l{2.2'}\ee
So, as in (\r{2.1'}), the aim of quantum perturbation theory is to
define the corrections to the phase $S_1(u_c)$

In terms of integrals the cancellation phenomena, shown in (\r{2.2}),
looks as follows:
\be
\R(E)=\sum_n\le|\f{1}{E-E_n-i\e}\ri|^2=
\sum_n\int^\infty_0 dT_+dT_-e^{-\e(T_++T_-)+i(E-E_n)(T_+-T_-)}
\l{b1'}\ee
To see the effect of cancellation let us introduce new time variables
$T$ and $\tau$:
\be
T_\pm=T\pm\tau.
\l{b1}\ee
The Jacobian of transformation gives: $0\leq
T\leq\infty$ and $-T\leq\tau\leq T$.  But in the integral (\r{b1'})
$T\sim(1/\e)\to\infty$
are essential at $\e\to0$.  By this reason we can put
$|\tau|\leq\infty$. In result,
\be
\e\R(E)=2\pi\e\int^\infty_0 dT e^{2\e T}\int^{+\infty}_{-\infty}
\f{d\tau}{\pi}e^{2i(E-E_n)\tau}
\l{b2}\ee
In the last integral all contributions, except for the case $E=E_n$, are 
cancelled.

Described  cancellation is not accidental, or approximate, being the
consequence of optical theorem, i.e. is the consequence of unitarity
condition. The $\d$-likeness of measure (\r{1.2}) has the same nature
as the $\d$-function in the r.h.s. of (\r{2.2}), i.e. the $\d$-like
measure will arise when the absorption part of amplitudes is
calculated. 

Note, we start from claculation of modulo squire of amplitudes 
since we know the path integral reprisantation for them. Then, using 
the unitarity condition we find the correct measure for imaginary part 
of the amplitude. 

\subsection{Functional $\d$-like measure}

We will use following path-integral representation for amplitude
\be
a(u_1,u_2;E)=i\int^\infty_0 dTe^{iET}\int_{u(0)=u_1}^{u(T)=u_2}Du
e^{iS_{C_+(T)}(u)},~Du=\prod_{t\in C_+(T)}\f{du(t)}{\sqrt{2\pi}},
\l{2.4}\ee
to calculate
\be
\R(E)=\int du_1du_2 \le|a(u_1,u_2;E)\ri|^2.
\l{2.6}\ee
The action $S_{C_+(T)}(u)$ is defined on the Mills complex time 
contour \C{mills}:
\be
C_\pm(T):~t\to t\pm i\e,~\e\to+0,~0\leq t\leq T
\l{2.5}\ee
Inserting $a(u_1,u_2;E)$ into (\r{2.6}) we find:
\be
\R(E)=\int^\infty_0 dT_+dT_-e^{iE(T_+-T_-)}
\int_{u_+(0)=u_-(0)}^{u_+(T_+)=u_-(T_-)}Du_+Du_-
e^{iS_{C_+(T_+)}(u_+)-iS_{C_-(T_-)}(u_-)}
\l{2.7}\ee
Note crucial for us the `closed-path' boundary conditions:
\be
u_+(0)=u_-(0),~u_+(T_+)=u_-(T_-).
\l{2.9}\ee

We will introduce new variables $T$ and $\tau$, see (\r{b1}). The 
integral over $\tau$ will be calculated perturbatively. In zero order 
over $\tau$ we would have from (\r{2.9}):
\be
u_+(0)=u_-(0),~u_+(T)=u_-(T).
\l{2.11}\ee
It should be underlined that this is unique solution of the boundary 
condition (\r{2.9}) which did not contradict to the quantum uncertainty 
principle (other solutions of (\r{2.9}) would involve constraints for 
time derivatives of coordinate).

If we introduce now new coordinates $u$ and $x$:
\be
u_\pm(t)=u(t)\pm x(t),
\l{2.12}\ee
then (\r{2.11}) gives:
\be
x(0)=x(T)=0
\l{2.13}\ee
and $u(0)$ and $u(T)$ are arbitrary. We will see that this `minimal' 
boundary condition is sufficient to define the integrals over $\tau$ 
and $u$. 

Let us expand the closed path action  
$$
S_{cl}(u\pm x;T\pm\tau)\equiv
(S_{C_+(T+\tau)}(u+x)-S_{C_-(T-\tau)}(u-x))
$$ 
over $\tau$:
\be
S_{cl}(u\pm x;T\pm\tau)=S_{cl}(u\pm x;T)-2\tau H_T(u)-
2\tilde{H}_T(u;\tau),
\l{2.14}\ee
where the Hamiltonian at the time moment $T$
\be
H_T(u)=-\f{\pa}{\pa T}S_{C_+(T)}(u).
\l{2.15}\ee
is $x$ independent because of (\r{2.13}). The remainder term  
$\tilde{H}_T(u;\tau)$ contains higher powers over $\tau$:
$$
\tilde{H}_T(u;\tau)=\sum^{\infty}_{n=1}\f{\tau^{2n+1}}{(2n+1)!}
\f{d^{2n}}{dT^{2n}}H_T(u).
$$
Therefore, the conditions (\r{2.13}) factorize $\tau$ and $x(t)$ 
dependence: the $x$ dependence is contained in the $\tau$ independent 
quantity $S_{cl}(u\pm x;T)$ only. So, we may construct the 
perturbation theory over $\tau$ and $x$ independently.  

Let us consider now expansion over $x$:
\be
S_{cl}(u\pm x;T)=S_{P(T)}(u)-
2{\rm Re}\int_{C_+(T)}dt x(t)\f{\d S_{C_+}(u)}{\d u(t)}-2\tilde{V}_T(u,x),
\l{2.17}\ee
where the first term in this decomposition is:
\be
S_{P(T)}(u)=(S_{C_+(T)}(u)-S_{C_-(T)}(u)).
\l{2.18}\ee
If the motion is periodic then $S_{P(T)}(u)$ is not equal to zero even 
on the real time axis \C{yaph}. In semiclassical approximation
$$
S_P(T)(u_c)=kS_1(u_c), k=0,1,...,
$$
and is $T$ independent. The reason of this conclusion is 
explained in Sec.3.3. As usual,
\be
2{\rm Re}\int_{C_+}dt=\int_{C_+}dt +\int_{C_-}dt
\l{2.21}\ee  
since for arbitrary analytic function $f(t\in C_+)=f^*(t\in C_-)$.

Following formal trick will be useful. We can write:
$$
e^{-2i\tilde{H}_T(u;\tau)}=\sum_n\f{\tau^n}{n!}K_n(u;T),
$$
where
$$
K_n(u;T)=\f{d^n}{d\tau_1^n}e^{-2iH_T(u;\tau_1)}|_{\tau_1=0}\equiv
\hat{\tau}_1^ne^{-2i\tilde{H}_T(u;\tau_1)}.
$$
On other hand,
$$
(2i\tau)^n=\f{d^n}{d\ve^n}e^{2i\ve\tau}|_{\ve=0}\equiv
\hat\ve^ne^{2i\ve\tau}.
$$
Therefore,
\be
e^{-2i\tilde{H}_T(u;\tau)}=\sum_n\f{(\h\tau_1\h\ve/2i)^n}{n!}
e^{2i\ve\tau}
e^{-2i\tilde{H}_T(u;\tau_1)}=
e^{-i\h\tau_1\h\ve/2}e^{2i\ve\tau}e^{-2i\tilde{H}_T(u;\tau_1)}.
\l{2.19}\ee
The expansion of the operator $e^{-i\h \tau_1\h \ve/2}$ will generate 
corresponding perturbation series.

The same operator can be introduced for expansion over the local 
quantity $x$:
\be
e^{-2i\tilde{V}_T(u,x)}=
e^{-\f{i}{2}{\rm Re}\int_{C_+}dt\h{j}(t)\h{x}_1(t)}
e^{2i{\rm Re}\int_{C_+}dtj(t)x(t)}e^{-2i\tilde{V}_T(u,x_1)}.
\l{2.20}\ee

Note, the eqs.(\r{2.19}), (\r{2.20}) linearise the arguments of
corresponding exponents. Then, using (\r{2.14}), (\r{2.17}) and
(\r{2.19}), (\r{2.20}) we find  that
\be
\R(E)=2\pi\int^\infty_0 dTe^{-i{\bf K(\ve\tau,jx)}}
\int DM(u)\d(E+\ve-H_T(u))
e^{iS_{P(T)}(u)}e^{-2i\tilde{H}_T(u;\tau)-2i\tilde{V}_T(u,x)},
\l{2.23}\ee
where expansion over the operator 
\be
{\bf K}=\f{1}{2}\le(\h\tau\h\ve+
{\rm Re}\int_{C_+}dt\h{j}(t)\h{x}(t)\ri)
\l{2.23'}\ee
gives the perturbation series. At the very end of calculations all 
auxiliary variables $\tau,~\e,~j$ and $x$ should be taken equal to 
zero.

The measure in (\r{2.23}) is defined as follows:
\be
DM(u)=\prod_t du\d\le(-\f{\d S(u)}{\d u}-j\ri)=
\prod_t du\d(\ddot{u}+v'(u)-j)
\l{2.26}\ee
and the $\d$-function is defined by the equality:
\be
\prod_t\d(\ddot{u}+v'(u)-j)=\int^{x(T)=0}_{x(0)=0}
\prod_t\f{dx}{\pi}
e^{2i{\rm Re}\int dtx(\ddot{u}+v'(u)-j)}.
\l{2.27}\ee
Argument of this $\d$-function did not contain the boundary 
values $u(0)$ and $u(T)$. But this  is not important since to 
solve the second order equation (\r{2.28})  two
constant of integration is necessary.

The exponent in (\r{2.27}) is equal to the sum:
${\rm Re}x~{\rm Re}(\ddot{u}+v'(u)-j)+ {\rm Im}x~{\rm
Im}(\ddot{u}+v'(u)-j)$, being defined on the complex  time contour. 
This means  that
\be
\prod_t\d(\ddot{u}+v'(u)-j)=\prod_{t\in C}\d({\rm Re}
\{\ddot{u}+v'(u)-j\})
\d(i{\rm Im}\{\ddot{u}+v'(u)-j\}),
\l{2.29}\ee
where $C=C_++C_-$. So, the measure (\r{2.26}) defines both the 
real and imaginary part of contributions.

By definition, $(\ddot{u}+v'(u)-j)$ is the total force, then the
product  $(\ddot{u}+v'(u)-j)x$ is the virtual work. In classical
mechanics this  work should be equal to zero, since the classical
motion is time reversible (d'Alembert). Then, noting that virtual
deviation is arbitrary, one finds the local condition:  
\be
\ddot{u}+v'(u)-j=0.   
\l{2.28}\ee  
when the motion is time reversible.

In quantum case the virtual work is not equal to zero (quantum
corrections shift the energy levels), but the integral over $x(t)$
gives the same result (\r{2.28}). We can conclude that the unitarity
condition  of quantum mechanics allows to $derive$ the d`Alembert's
variational principle  of classics mechanics \C{yaph}, see also 
\C{fok}.

\subsection{Properties of theory with $\d$-like measure}

The eq.(\r{2.28}) should be solved expanding over $j(t)$:
\be
u_j(t)=u_c(t)+\int dt'G(t,t';u_c)j(t')+({\rm higher~powers~of}~j)
\l{2.30}\ee
where $u_c$ is the solution of homogeneous equation:
\be
\ddot{u}+v'(u)=0
\l{2.31}\ee
and $G(t,t';u_c)$ is the Green function:
\be
(\pa^2_t+v''(u_c))G(t,t';u_c)=\d(t-t').
\l{2.32}\ee

The eq.(\r{2.31}) have in our case the trivial constant solution
\be
u_0:~\dot{u}_0=0,~v'(u_0)=0
\l{2.33}\ee
and nontrivial one 
\be
u_c=u_c(t):~\dot{u}_c(t)\neq 0,~\ddot{u}_c+v'(u_c)\equiv 0.
\l{2.34}\ee
Because of definition of the $\d$-function and since there is 
not any special restriction on the contributions both one should be 
taken into account:
\be
\R(E)=\R_0(E)+\R_c(E).
\l{2.35}\ee

This means that one should sum over all possible topological classes
of trajectory, if the single class is unable to cover all phase
space. Each class of trajectories belongs to restricted domain of
phase space: $\O=W^0\t W^c$ in our case. Each sub-domain $W^i$ is
restricted by the bifurcation lines
\C{smale}. This means that $\R_0$ can not be achieved
by analytical continuation of $\R_c$ (for instance, taking  $E=0$ in
$\R_c$ for the semiclassical  approximation).

It  is evident, one should leave in the sum (\r{2.35}) the term with
higher volume $V_{W^i}$, $i=0,c$. This is just the domain of  $u_c\in
W^c$ trajectory,  and one can put out $\R_0$ since the sub-domain of
$u_c^0\in W^0$ is  the point \C{yaph, smale}. Indeed, it will be shown
that $\R_c\sim V_{tr}=\infty$, where $V_{tr}$ is the volume of time
translations mode (zero frequency mode). At the same time, $\R_0\sim
O(1)$. Therefore, iff the time translation invariance is unbroken,
one can say that $\R_0$ is realised on the measure $\sim
O(1)/V_{tr}=0$.

The ability to classify contributions by the trajectory topology
classes becomes  possible since  there is not in (\r{2.35}) the
$u_c^0$ and $u_c$ interference term. This is evident consequence of
the orthogonality of corresponding Hilbert spaces. Therefore, the
choice of solution means choice of corresponding vacuum.

\section{WKB perturbation theory}\0

It can be shown that (\r{2.23}) restores ordinary WKB expansion. The
first step of this calculations is to find the solution of
inhomogeneous equation (\r{2.28}), Sec.3.1. Then we may find that this
perturbation theory counts positive powers of $g$, Sec.3.2. At the end
the zero frequency modes problem will be discussed.

\subsection{Tree decomposition}

Let us consider the tree decomposition (\r{2.30}) more carefully for 
the potential
\be
v(u;g)=\f{1}{2}w_0^2u^2+\f{1}{4}gu^4.   
\l{2.36}\ee
It is evident:
\be
v(u;g)=\f{1}{g}v(g^{1/2}u)
\l{2.37}\ee
The decomposition (\r{2.30}) can be written in the form:
\be
u_j(t)=u_c(t)+\sum_{n=1}^\infty\f{1}{n!}\int \prod_{k=1}^n
\le\{dt_ij(t_i)\ri\}G_n(t,t_1,....,t_n;u_c)
\l{2.38}\ee
It easy to show that the $n$-point Green function \C{jack}
\be
G_n=O(g^{(n-1)/2}).
\l{2.39}\ee

Indeed, inserting (\r{2.38}) into the equation:
\be
\ddot{u}+\o_0^2u+gu^3=j
\l{2.40}\ee
we find:
\be
(\pa^2_t+\o_0^2+3gu_c^2)G_1(t,t_1;u_c)=\d(t-t_1)
\l{2.41}\ee
The  operator $(\pa^2_t+\o^2+3gu_c^2)$ is $g$ independent since
$u_c=O(1/g^{1/2})$ and, therefore, $G_1=O(g^0)$.

Note  also, the  operator  ($\pa^2_t+\o^2+3gu_c^2$) is  translationally
noninvariant.  By  this reason  considered  perturbation theory  is
sufficiently  complicated  so that  only  first  corrections has  been
computed till now.

The equation for $G_2$ have the form:
\be
(\pa^2_t+\o^2+3gu_c^2)G_2(t,t_1,t_2;u_c)+6gu_cG_1(t,t_1;u_c)
G_1(t,t_2;u_c)=0.
\l{2.42}\ee
Therefore, in accordance with (\r{2.39}), 
$G_2=O(g^{1/2})$. In result, the analysis of higher orders  over $j$
would justify (\r{2.39}).

The interactions generating functional $V_T$ computed for the case 
(\r{2.36}) has the form:
\be
\tilde V_T(u,x)=2g{\rm Re}\int_{C_+}dt x^3(t)u(t)+O(\e),
\l{2.43}\ee
where the $O(\e)$ term is proportional to the imaginary part of $S_{cl}$.

The operator $\bf{K}$ is linear over $\h{x}=\d/\d x$. Therefore, action of 
$\exp\{-i\bf{K}\}$ will give:
\be
\R_c\sim :e^{-2i\tilde{\bf V}_T(u,\h{j}/2i)}:e^{iS_{P(T)}(u)}
e^{-2i\tilde{H}_T(u;\tau)}\d(E+\e-H_T(u_j)),
\l{2.44}\ee
where the colons prescribe normal product, when the operator should
stay to the left of all functions on which it may act, and the
unimportant for present analyses integrations were not mentioned.. The
expansion of  the operator exponent gives the perturbation series:
\be
\R_c\sim\sum_n\f{(-2i)^n}{n!}:\tilde{\bf V}^n_T(u_j,\h{j}/2i):
e^{iS_{P(T)}(u_j)}
e^{-2i\tilde{H}_T(u_j;\tau)}\d(E+\e-H
_T(u_j)).
\l{2.44a}\ee

Let us consider the self-interaction part for the beginning. This
means that the shifting energy levels \C{dash, yaph} renormalisation
of $S_P$ and $\tilde{H}_T$ is not considered. In other words, we omit
the action of operators $\prod\h{j}(t_i)$ on
$\exp\{iS_{P}(u_j)-2i\tilde{H}_T(u_j; \tau)\}$:
$$
\R_c\sim \d(E+\e-H_T(u_j))e^{iS_{P(T)}(u_c)}
e^{-2i\tilde{H}_T(u_c;\tau)}\sum_n\f{(-21)^n}{n!}
:\tilde{\bf V}^n_T(u_j,\h{j}/2i):. 
$$
Then   the lowest order
contribution is $\sim\tilde{\bf V}$. So, in the first order we find:
\be
\sim\h{j}^3u_j\sim G_3 =O(g^2),
\l{2.45}\ee
where the estimation (\r{2.39}) was used and the prescription that the 
auxiliary variable $j$ should be taken equal to zero was taken into account.
In the second order $\sim \tilde{\bf V}_T^2=O(\h{j}^6)$ contribution 
have following 
order over $g$:
\be
\sim g^2\h{j}^6u_j^2=O(g^4),
\l{2.46}\ee
and so on. 

In result, one can find that the $n$-th order in expansion of 
$\exp\{-i\bf K\}$ gives $O(g^{2n})$ expansion if the self-interactions 
only are included in $\R(E)$.

As follows from decomposition (\r{2.38}) and estimation (\r{2.39}) the
action of operator $\h j$ on $u_j$ gives coefficient $\sim
g^{1/2}$. The renormalisation of $S_P$ and $\tilde H_T$ start from
$\sim 1/g$ terms, but higher orders would contain the positive powers
of $g$ since they are produced by the actions of $\prod \h j(t_i)$.

\subsection{Connection with WKB expansion}

One can show another argument that considered above perturbation
theory is nothing  new but is the ordinary expansion around $u_c$
developed in  early publications \C{jack, jew}. Let us use for this
purpose  the substitution:
\be
u(t)\to u_c(t)+u(t) 
\l{3.1}\ee
Then
\ba
\R_c(E)=2\pi\int^\infty_0 dT
e^{-i{\bf K(\ve\tau,jx)}}
\int DM(u_c,u)\d(E+\ve-H_T(u_c+u))\t
\n\\\t
e^{iS_{P(T)}(u_c+u)}
e^{-2i\tilde{H}_T(u_c+u;\tau)-2i\tilde{V}_T(u_c+u,x)},
\l{3.2}\ea
where
\be
DM(u_c,u)=\prod_t du\d\le(\f{\d S(u_c+u)}{\d u}+j\ri)
\l{3.2'}\ee

We should take into account that $u_c$ depends on the integration
constants $\x$ and  $\eta$. Therefore, if $(\x,\eta)$ form the
manifold $W^c$, as was mentioned in Sec.2.3, one should sum over all
solutions $u_c\in W^c$, see Sec.3.3.

We want to show now that (\r{3.2}) may be reduced to the product of
two  path integrals. Indeed, using (\r{2.27}) and (\r{2.14}),
(\r{2.17}) we find  from (\r{3.2}) that
\ba
\R_c(E)=2\int^\infty_0 dT\int^{+\infty}_{-\infty}d\tau' 
e^{-i\bf K}\int DuDx'
e^{2i(E+\ve-H_T(u_c+u))\tau'}e^{2iH_T(u_c+u)\tau}
\t\n\\\t 
e^{S_{cl}(u_c+u\pm x';T\pm\tau')}e^{2i{\rm Re}\int dt x(\pa S(u_c+u)/\pa u)}
e^{-2i{\rm Re}\int dt x\{(\pa S(u_c+u)/\pa u)+j\}}.
\l{3.3}\ea
The action of operator $\exp\{-i\bf K\}$ leads to substitutions:
$
x\to x',~\tau\to\tau'$ and $\ve\to0,~j\to0.
$
Taking this into account we find:
\be
\R_c(E)=\le|\int^\infty_0 dTe^{iET}
\int Du e^{iS_{C_+}(u_c+u)}\ri|^2,
\l{3.7}\ee
where the functional integral should be calculated perturbatively 
over $u$. Note, calculation of amplitudes is useful since eliminates 
the doubling of degrees of freedom.

\subsection{Zero modes}

The defined by eq.(\r{3.7}) $\R_c(E)$ stay undefined till the procedure
of summation over all $u_c\in W$ is not formulated. Following to the
equality:
$$
\sum_{\{u_c\}}=\int_W d\x d\eta \s(u;\x,\eta)
$$ 
we should define the density $\s(u;\x,\eta)$ of states in the
domain $(\x,\x+d\x;\eta,\eta+d\eta)$. The Faddeev-Popov $ansatz$ is
used for this purpose \C{jew}.

By definition, $(\x,\eta)$ are the constants of integration and they may
be chosen  arbitrarily. For example, wee may take $(\x,\eta)$ as the
initial  coordinate and momentum of particle on the trajectory
$u_c$. 

But the  `field-theoretical' definitions would be much more useful for
us, see Sec.4. One  may note that the dependence on $(\x, \eta)$
indicates the symmetry breaking. Then $\eta$ may be taken as the
generator $J$ of broken symmetry and $\x$ as the canonically conjugate
to  it coordinate $\Th$. It will be important for us that $(\x, \eta)$
define the solution $u_c$ unambiguously. In other words, we will use
the ordinary mechanical statement \C{smale} that $(\x,\eta)$ form a
manifold $W^c$ and $u_c$ belongs to it $completely$. So, we would assume
that the equations: 
\be  
\x=\Th(u_c,\dot{u_c}),~\eta =J(u_c,\dot{u_c})
\l{3.4}\ee  
define the integration constants of $u_c$ unambiguously.
 
Then, to define the density $\s$, we may insert into the initial
representation (\r{2.23}) the unite (Faddeev-Popov $ansatz$):
\be
1=\int_W \prod_t d\x d\eta \d(\x-\Th(u,\dot{u}))\d(\eta-J(u,\dot{u})
\l{3.8}\ee
Note, by definition $\eta$ should coincide with conserved generator. 
But nevertheless we consider $\eta=\eta(t)$ and the same for $\x$. 
This assumption is necessary since the quantum case is considered. 

We can change order of integration and integrate firstly over $u$
using  the $\d$-function of the measure $DM$. Lagrange equation
(\r{2.31}) should be solved taking into account the constraints
(\r{3.4}):
\be
\R_c(E)=2\pi\int_W d\x(0) \int^\infty_0 dTe^{-i{\bf K(\ve\tau,jx)}}
\int DM_c(u)\d(E+\ve-H_T(u))
e^{iS_{P(T)}(u)}e^{-2i\tilde{H}_T(u;\tau)-2i\tilde{V}_T(u,x)},
\l{3.5}\ee
where $\x(0)$ is the initial phase and the constraint measure
\be
DM_c(u)=\prod_t du\d\le(\f{\d S(u)}{\d u}+j\ri)
 \d\le(\x_0-\Th(u,\dot{u})\ri)
\l{3.5'}\ee
was introduced.  In our problem the value of $J$ is restricted by
$\d(E+\ve-H_T(u))$. It was used in (\r{3.5}) that $DM_c$ is $\x$
independent since Lagrange equation is invariant against $\x$
variations and $j=j(t)$ is the auxiliary variable. This means that
$\R_c(E)$ defined in (\r{3.5}) is proportional to the volume
$$V_{tr}=\int d\x(0)$$ of the time translation mode.
   
It is important here to trace on the following question. One can note
that (\r{3.5}) gives $\R_c\sim V_{tr}^1$. On other hand, as follows from
(\r{3.7}), one may expect $\R_c\sim V_{tr}^2$. It is evident that this
discrepancy is the consequence of loaded into formalism condition of
the orthogonality of Hilbert spaces, see Sec.2.3.

Remembering definition of $\R$ as the squire of amplitudes, we may
insert the Faddeev-Popov's unite defined on the whole time contour
$C=C_++C_-$, see (\r{2.29}), to take into account the input condition that the
trajectories $u_+(t\in C_+)$ and $u_-(t\in C_-)$ are absolutely
independent. This means that, generally speaking, the boundary
conditions for this trajectories should not coincide and, therefore,
if we introduce $\x(t\in C_\pm)|_{t=0}\equiv\x_\pm$, one should have
in mind that, generally speaking, $\x_+\neq\x_-$. Then integration
over $\x_+$ and $\x_-$ should be performed independently.

But we have considered the closed-path contributions, see (\r{2.11}). This
gives restriction for the $u_\pm$ boundary properties. Taking into
account (\r{2.13}), we can find, considering the periodic orbits, that
\be
\x_+=\x_-\pm kP_1(E),~k=0,1,2,...,
\l{3.9}\ee
where $P_1(E)$ is the elementary period. Just this solution leads to
$S_P\neq0$ and the necessary summation over $k$ gives the energy
levels quantisation condition (\r{2.2'}), see also \C{dash, yaph}.

\section{Mapping on the cotangent manifold}\0

The necessity to search a new form of the perturbation theory is
caused by extremal complexity of the WKB perturbation theory described
above.

The quantum nature of collective variables $(\x,\eta)$ was mentioned
previously by many authors \C{jack, fad}. We would like continue this
idea considering them as a new quantum variables. For this purpose we
will use the $\d$-like definition of measure, the definition of the
interactions generating functional $\tilde{V}_T$ and the perturbations
generating functional $\exp\{\bf K\}$, to count the possible
excitations of the field $u_c\in W^c$, see Sec.4.1. In Sec.4.2 we will
show the structure of new perturbation theory.

\subsection{Procedure of mapping}

Let us return to the Faddeev-Popov unite
\be
1=\int D\x D\eta\prod_t \d(\x-\P(u,\dot{u}))\d(\eta -J(u,\dot{u}))
\l{4.1}\ee
It is assumed, as was offered in Sec.3.3, 
$$
\prod_t\equiv\prod_{t\in C=C_++C_-}.
$$

The first order formalism will be useful for us . Corresponding 
measure
\be
DM(u,p)=\prod_t du dp \d\le(\dot{u}-\f{\pa H_j(u,p)}{\pa p}\ri)
\d\le(\dot{p}+\f{\pa H_j(u,p)}{\pa u}\ri),
\l{4.2}\ee
where the total Hamiltonian
\be
H_j(u,p)=\f{1}{2}p^2+v(u)-ju 
\l{4.3}\ee
includes the energy of quantum excitations $ju$. It is evident that 
the integration over $p$ gives incident measure (\r{2.26}).

Inserting (\r{4.1}) into the functional integral with measure
(\r{4.2})  we find that we have four equations for $u$ and $p$: 
\be  \dot{u}=\f{\pa H_j(u,p)}{\pa
p},~\dot{p}=-\f{\pa H_j(u,p)}{\pa u} 
\l{4.4}\ee 
and 
\be
\x(t)=\P(u,\dot{u}),~\eta(t)=J(u,\dot{u}). 
\l{4.5}\ee 
In previous section the first pare of equations (\r{4.4}) was used to
calculate the functional integral.

But now we would like use second one (\r{4.5}). It is possible iff
$u_c$ belongs to the space $W^c$ completely and $W^c$ is a manifold. This
condition means that the  eqs.(\r{4.5}) have unique solution $(u_c,
p_c)$ and this solution  transform (\r{4.4}) into identity at least at
$j=0$. Let $u_c(\x,\eta)$ and $p_c(\x,\eta)$ are the solutions of
(\r{4.5}).  One can recognise in our description the ordinary canonical
transformation (\r{1.4}), i.e. it defines the cotangent foliation
$W^c=T^*\O$. But eqs.(\r{4.4}) and (\r{4.5}) should be solved simultaneously. So,
inserting $u_c,p_c$ into (\r{4.4}) we should use the `excited' by $j$
solutions $\x_j(t)$ and $\eta_j(t)$.  

So, we wish to adopt the statement that the random (Gaussian) walk,
induced by the same operator $\exp\{-i\bf K\}$, covers both
$W^c=(\x,\eta)$ and $\O=(u,p)$ spaces densely. By this reason one may
choose one of them arbitrarily.

The corresponding Jacobian of transformation $\D$ is $\d$-like:
\be
\D=\prod_t
\d(\dot{u}_c-\f{\pa H_j(u_c,p_c)}{\pa p_c})
\d(\dot{p}_c-\f{\pa H_j(u_c,p_c)}{\pa u_c}),
\l{4.6}\ee
and 
\be
det^{-1}(u_c,p_c)=\int\prod_t du dp \d(\x-\P(u,\dot{u}))
\d(\eta -J(u,\dot{u}))=1
\l{4.7}\ee
since the transformation is canonical. This allows to diagonalise 
$\D$ and mapping into the $W^c$ space leads to following path 
integral representation:
\be
\R_c(E)=2\pi\int dT e^{-i\bf K}\int DM(\x,\eta)\d(E+\e-h(\x,\eta;T))
e^{iS_{P(T)}(u_c)}e^{-2i\tilde{h}(u_c;\tau,T)-2i\tilde{V}_T(u_c,x)},
\l{4.8}\ee  
where the measure
\be
DM(\x,\eta)=\prod_t d\x d\eta 
\d\le(\dot{\x}-\f{\pa h_j(\x,\eta)}{\pa \eta}\ri)
\d\le(\dot{\eta}+\f{\pa h_j(\x,\eta)}{\pa \x}\ri)
\l{4.9}\ee
and $h_j$ is the transformed Hamiltonian:
\be
h_j(\x,\eta)=h(\eta)-ju_c(\x,\eta).
\l{4.10}\ee

In result of mapping the problem of calculation of functional integral 
was reduced to solution of equations:
\be 
\dot{\x}=\f{\pa h_j(\x,\eta)}{\pa\eta}=
\o(\eta)-j\f{u_c(\x,\eta)}{\pa\eta},
~\dot{\eta}=-\f{\pa h_j(\x,\eta)}{\pa\x}=j\f{u_c(\x,\eta)}{\pa\x},
\l{4.11}\ee
where one can choose, for example, 
\be
\o(\eta)=\f{\pa h(\eta)}{\pa\eta}=1.
\l{4.12}\ee
This means that in this case 
\be
\eta=H(u,p),~\x=\int^u \f{dy}{\sqrt{2(H-v(y))}}.
\l{4.13}\ee
It is evident that the solution of this equations gives $u_c(\x,\eta)$
and $p_c(\x,\eta)$ unambiguously.

\subsection{Structure of transformed perturbation theory}

We want to show now that $\R(E)$, defined in (\r{4.8}), has the strong 
coupling expansion. Let us start for this purpose from the 
`tree decomposition' of the equations (\r{4.11}):
\be
\dot{\x}=\f{\pa h_j(\x,\eta)}{\pa\eta}=
1-j\f{u_c(\x,\eta)}{\pa\eta},
~\dot{\eta}=-\f{\pa h_j(\x,\eta)}{\pa\x}=j\f{u_c(\x,\eta)}{\pa\x}.
\l{v.1}\ee
We will consider following decomposition of the solutions $\x_j$ 
and $\eta_j$:
\ba
\x_j(t)=\x_0(t)+\sum_n\f{1}{n!}\int\prod\{dt_ij(t_i)\}
\x_n(t;t_1,...,t_n),
\n\\
\eta_j(t)=\eta_0(t)+\sum_n\f{1}{n!}\int\prod\{dt_ij(t_i)\}
\eta_n(t;t_1,...,t_n).
\l{v.2}\ea
Inserting (\r{v.2}) into the (\r{v.1}) we  find equation for the  
$n$-point Green functions $\x_n(t;t_1,...,t_n)$ and 
$\eta_n(t;t_1,...,t_n)$. It can be shown:
\be
\x_n=O(g^{-n/2}),~\eta_n=O(g^{-n/2}).
\l{v.3}\ee
Indeed, in zero order over $j$ we have:
\be
\x_0=\x(0)+t,~\eta_0=\eta(0)
\l{v.4}\ee
since $W^c$ is the homogeneous and isotropic manifold.
Then, in the first order over $j$:
\be
\dot{\x}_1(t;t_1)=\d(t-t_1)\f{\pa u_c(\x_0(t),\eta_0)}{\pa\eta_0}=
O(g^{-1/2}),~
\dot{\eta}_1(t;t_1)=\d(t-t_1)\f{\pa u_c(\x_0(t),\eta_0)}{\pa\x_0}=
O(g^{-1/2})
\l{v.5}\ee
since the derivatives of $u_c$ are unable to change the $g$ dependence.
In second order we have the equations:
\ba
\dot{\x}_2(t;t_1,t_2)=\d(t-t_1)\le\{\x_1(t;t_2)
\f{\pa u_c(\x_0(t),\eta_0)}{\pa\eta_0\pa\x_0}+
\eta_1(t;t_2)\f{\pa u_c(\x_0(t),\eta_0)}{\pa\eta_0\pa\eta_0}\ri\}=
O(g^{-1}),
\n\\
\dot{\x}_2(t;t_1,t_2)=\d(t-t_1)\le\{\x_1(t;t_2)
\f{\pa u_c(\x_0(t),\eta_0)}{\pa\x_0\pa\x_0}+
\eta_1(t;t_2)\f{\pa u_c(\x_0(t),\eta_0)}{\pa\x_0\pa\eta_0}\ri\}=
O(g^{-1})
\l{v.6}\ea
And so on. So, each power of $\h j$ adds $g^{-1/2}$. This proves  
the estimation (\r{v.3}). It is important to note that eqs.(\r{v.6}) 
are trivially integrable. Therefore, we can calculate $(\x_n,\eta_n)$ 
for arbitrary $n$. 

The operator $\bf K$ is linear  over $\h{x}$. So, the result 
of its action gives the normal ordered structure:
\be
:e^{-2i\tilde{\bf V}_T(u_c,{\h j}/2i)}: e^{iS_{P(T)}(u_c)}
e^{-2i\tilde{H}_T(u_c;\tau)}\d(E+\e-h(\x_j,\eta_j;T)),
\l{5.2}\ee 
and at the very end of calculations one should take the auxiliary 
variables $x$ equal to zero.

Let us consider once more the $gu^4$ theory. Then,  
$$
\tilde{V}_T(u_c,\h{j}/2i)=O(\h{j}^3),
$$ 
Therefore, leaving the  self-interaction parts only, in the lowest 
order we would have the contribution
\be
\sim \tilde{V}_T(u_c,\h{j}/2i)\sim g\h{j}^3
u_c=O(1/g)
\l{5.4}\ee
where (\r{v.3}) was used. So, the lowest order of new perturbation 
theory  is $\sim 1/g$.  In result, the $n$-th order is $\sim 
\tilde{V}_T(u_c,{\h j}/2i)^n\sim1/g^n$.

The action of $\tilde{V}_T^n(u_c,\h{j}/2i)$ on $e^{iS_{P(T)}(u_c)}
e^{-2i\tilde{H}_T(u_c;\tau)}\d(E+\e-h(\x_j,\eta_j;T))$ did not alter this conclusion since the derivative of $u_c$ can not change the $g$ dependence.

\section{Conclusion}\0 

We conclude this paper by notation that it is impossible the
transformed theory reduce to amplitude representation. Indeed, let us
return to (\r{4.8}) and use the Fourier definition of the
$\d$-functions:
\ba
\R_c(E)=\int^\infty_0 dT\int^{+\infty}_{-\infty} d\tau'e^{-i\bf K}
\int D\x D\eta Dx_\x Dx_\eta e^{2i(E+\e-h(\eta;T))\tau'}
\t\n\\\t
e^{-2i{\rm Re}\int_{C_+}dtx_\x\{\dot{\x}-
{\pa h_j(\x,\eta)}/{\pa \eta}\}}
e^{-2i{\rm Re}\int_{C_+}dtx_\eta\{\dot{\eta}+
{\pa h_j(\x,\eta)}/{\pa \x}\}}
e^{iS_{P(T)}(u_c)}e^{-2i\tilde{h}(u_c;\tau,T)-2i\tilde{V}_T(u_c,x)},
\l{vi.1}\ea 
where, see (\r{2.14}), 
\be
-2\tilde{h}(u_c;\tau,T)=S_{cl}(u_c\pm x;T\pm\tau)-S_{cl}(u_c\pm x;T)+
2\tau h(\eta).
\l{vi.2}\ee
Using this definition, and remembering that the action of operator
$\exp\{-i\h{e}\h{\tau}/2\}$ gives $\tau=\tau'$ and $\e=0$, we find:
\ba
\R_c(E)=\int^\infty_0 dT\int^{+\infty}_{-\infty} d\tau e^{2iE\tau}
e^{-i{\rm Re}\int_{C_+}dt\h{j}\h{x}/2}
\int D\x D\eta Dx_\x Dx_\eta 
e^{iS_{cl}(u_c\pm x;T\pm\tau)-iS_{cl}(u_c\pm x;T)}
\t\n\\\t
e^{-2i{\rm Re}\int_{C_+}dtx_\x\d S(u_c)/\d \eta}
e^{2i{\rm Re}\int_{C_+}dtx_\eta\d S(u_c)/\d \x}
e^{iS_{P(T)}(u_c)}e^{-2i\tilde{V}_T(u_c,x)},
\l{vi.3}\ea 
if the transformed action
$$
S(u_c)=\int dt\{\eta\dot\x -h(\eta)\}.
$$ 

Action of the perturbation generating 
operator gives:
\ba
\R_c(E)=\int^\infty_0 dT\int^{+\infty}_{-\infty} d\tau e^{2iE\tau}
\int D\x D\eta Dx_\x Dx_\eta e^{iS_{cl}(u_c\pm x_c;T\pm\tau)}
\t\n\\\t
e^{-2i{\rm Re}\int_{C_+}dt\{x_\x(\d S(u_c)/\d \eta)-
x_\eta(\d S(u_c)/\d \x) \}}
e^{-2i{\rm Re}\int_{C_+}dtx_c(\d S(u_c)/\d u_c)},
\l{vi.4}\ea
if (\r{2.17}) is used and, using the local coordinates of the $W$ space, 
\be
x_c=x_\x\f{\pa u_c}{\pa\eta}-x_\eta\f{\pa u_c}{\pa\x}=
\d u_c\wedge\d p_c
\l{vi.5}\ee
Now, if
\be
\f{\d S(u_c)}{\d\x}=\f{\pa u_c}{\pa\x}\f{\d S(u_c)}{\d u_c},~
\f{\d S(u_c)}{\d\eta}=\f{\pa u_c}{\pa\eta}\f{\d S(u_c)}{\d u_c},
\l{vi.6}\ee
then we can write:
\be
\R_c(E)=\int^\infty_0 dT\int^{+\infty}_{-\infty} d\tau e^{2iE\tau}
\int D\x D\eta Dx_\x Dx_\eta e^{iS_{cl}(u_c\pm x_c;T\pm\tau)}.
\l{vi.7}\ee 
The quantities $(x_\x,x_\eta)$ and $(\x,\eta)$ have different
meaning. First ones are the virtual variation of the `field' $u$ along
the corresponding axis of $W^c$ space, and the integrals over them
should be calculated perturbatively, but last ones are the axis of the
$W^c=T^*\O$ phase space. The closed path action
\be
S_{cl}(u_c\pm x_c;T\pm\tau)=
S_{C_+(T+\tau)}(u_c(\x,\eta)+x_c(\x,\eta;x_\x,x_\eta))-
S_{C_-(T-\tau)}(u_c(\x,\eta)-x_c(\x,\eta;x_\x,x_\eta)).
\l{vi.8}\ee
It is evident from (\r{vi.7}) the transformed representation can not
be written in the factorized form of product of two amplitudes.  

We interpret this conclusion as impossibility of the canonical
transformations in the path integrals (\r{2.4}) since on the cotangent
manifolds the quantum excitations induce the phase space flows in which
all degrees of freedom are mixed. 

In traditional terms this means the problem of time ordering of
nonlinear operators. Our success is based on the observation that the
unitarity condition unambiguously defines the perturbation theory in
the (`linear') representation, where we may disentangle all time
orderings. Fixing this procedure in the structure of $DM$, $\b K$ and
$\tilde V_T$ one can do arbitrary transformations. But, the payment
for this success is necessity to work in terms of less habitual
absorption part of amplitude and, by this reason, one should be
careful interpreting our perturbation theory as a general, see \C{hep}.

But, in conclusion, quantising the nonlinear waves our strong coupling 
perturbation theory seems much more attractive since we can perform the 
calculation in this theory up to the end, choosing $W^c$ as the  
homogeneous and isotropic space. 
\vskip 0.5cm
{\bf Acknowledgement}\\
We would like to thank V.Kadyshevsky for interest to described 
perturbation theory.

\end{document}